\documentclass[aps,prb,reprint,superscriptaddress]{revtex4-2}
	\usepackage[utf8]{inputenc}
	
	\usepackage[hidelinks]{hyperref}
	
	\usepackage{hyperref}
	\usepackage{color}
	\hypersetup{colorlinks=true}
	\usepackage{graphicx}
	\usepackage{bm}
	\usepackage{amsmath}
	\usepackage{amssymb}
	\usepackage{xspace}
	\usepackage{mathtools}
	\usepackage{physics}
	\usepackage{siunitx}
	\usepackage{chemformula}
	\usepackage{setspace}

	\usepackage[textsize=footnotesize]{todonotes}

	\newcommand{\blockcomment}[1]{}
	
	\graphicspath{{./figures/}}
	\begin{document}
		
		\title{Charge conservation in spin torque oscillators leads to a self-induced torque}

		\author{Pieter M. Gunnink}
		\altaffiliation[Currently at ]{Institute of Physics, Johannes Gutenberg-University Mainz, Staudingerweg 7, Mainz 55128, Germany}
	\email{pgunnink@uni-mainz.de}
		\affiliation{Institute for Theoretical Physics and Center for Extreme Matter and Emergent Phenomena, Utrecht University, Leuvenlaan 4, 3584 CE Utrecht, The Netherlands}
	
		\author{Tim Ludwig}
		\affiliation{Institute for Theoretical Physics and Center for Extreme Matter and Emergent Phenomena, Utrecht University, Leuvenlaan 4, 3584 CE Utrecht, The Netherlands}
		
		\author{Rembert A. Duine}
		\affiliation{Institute for Theoretical Physics and Center for Extreme Matter and Emergent Phenomena, Utrecht University, Leuvenlaan 4, 3584 CE Utrecht, The Netherlands}
		\affiliation{Department of Applied Physics, Eindhoven University of Technology, P.O. Box 513, 5600 MB Eindhoven, The Netherlands}
		
		\date{\today}
		\begin{abstract}
		Spin torque oscillators are conventionally described by the Landau-Lifshitz-Gilbert-Slonczewski (LLGS) equation. However, at the onset of oscillations, the predictions of the conventional LLGS equation differ qualitatively from experimental results and thus appear to be incomplete. In this work we show that taking charge conservation into account leads to a previously-overlooked self-induced torque, which modifies the LLGS equation. We show that the self-induced torque originates from the pumping current that a precessing magnetization drives through a magnetic tunnel junction. To illustrate the importance of the self-induced torque, we consider an in-plane magnetized nanopillar, where it gives clear qualitative corrections to the conventional LLGS description.
	
		\end{abstract}
		\maketitle

	\section{Introduction}
	\label{sec:intro}
	The conventional Slonczewski spin-transfer torque \cite{slonczewskiCurrentdrivenExcitationMagnetic1996,ralphSpinTransferTorques2008} provides the basis for the description of steady-state precessions in spin torque oscillators (STO), a versatile functional element of spintronics \cite{slavinNonlinearAutoOscillatorTheory2009,kimChapterFourSpinTorque2012,chenSpinTorqueSpinHallNanoOscillators2016}. The corresponding Landau-Lifshitz-Gilbert-Slonczewski (LLGS) equation \cite{slonczewskiCurrentdrivenExcitationMagnetic1996,ralphSpinTransferTorques2008}, however, does not accurately predict the observed precession frequency  \cite{kiselevMicrowaveOscillationsNanomagnet2003,berkovMagnetizationPrecessionDue2005,xiaoMacrospinModelsSpin2005,tiberkevichNonlinearPhenomenologicalModel2007,berkovSpintorqueDrivenMagnetization2008}. A commonly employed solution to address the discrepancy between prediction and observation is to generalize the Gilbert damping by phenomenologically including nonlinear dissipation terms \cite{tiberkevichNonlinearPhenomenologicalModel2007,slavinNonlinearAutoOscillatorTheory2009}. Micromagnetic simulations that go beyond the single-domain approximation indicate that inhomogeneities in the magnetization could also play an important role \cite{miltatSpinTransferInhomogeneous2001,liMagnetizationDynamicsSpintransfer2003,berkovMagnetizationPrecessionDue2005}. Others have shown that additional corrections to the Slonczewski spin-transfer torque arise due to an interplay of spin conservation and the dynamics in adjacent layers \cite{tserkovnyakDynamicStiffnessSpin2003a,taniguchiCriticalCurrentSpintransfertorquedriven2008,chibaMagnetizationDampingNoncollinear2015,taniguchiSpincurrentDrivenSpontaneous2018}.

	In this work we show that imposing charge conservation in spin torque oscillators gives rise to a previously-overlooked self-induced torque, modifying the LLGS equation. This self-induced torque leads to qualitatively important corrections to the precession frequency and power and should thus be included in the description of the magnetization dynamics in spin torque oscillators. Given the fact that charge conservation is not easily violated, we believe the self-induced torque in this work to be of importance in a wide range of systems.

	\begin{figure}[b]
		
		\includegraphics[width=\columnwidth]{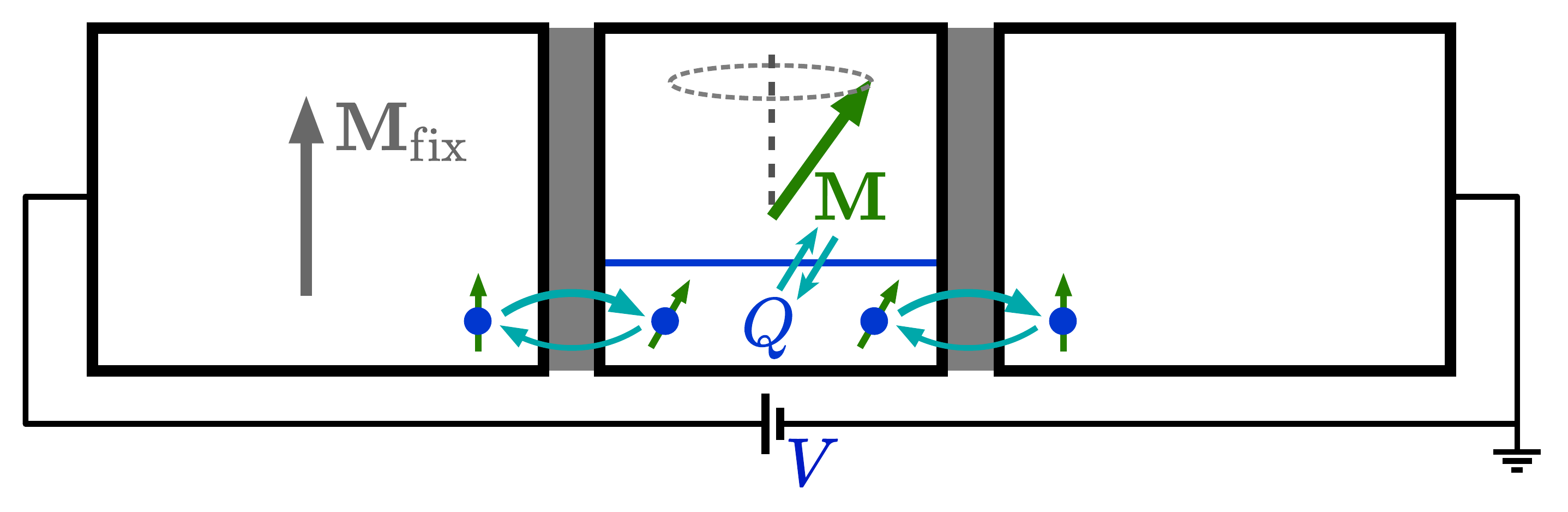}
		
		\caption{We consider a metallic nanomagnet with dynamical magnetization that is tunnel-coupled to a ferromagnetic lead (left) and a normal metal lead (right). The ferromagnetic lead has a fixed magnetization $\bm M_\mathrm{fix}$ that is parallel to the precession axis of the dynamical magnetization $\bm M$. A voltage $V$ is applied across the whole system. As shown in the main text, an interplay emerges between the nanomagnet's magnetization $\bm M$ and its charge $Q$. It is the result of charge conservation, in combination with the charge pumping caused by the precessing magnetization and the Slonczewski spin-transfer torque caused by the voltage drop across the magnetic tunnel junction. \label{fig:system}}
	\end{figure}

	We consider a magnetic double tunnel junction as shown in Fig.~\ref{fig:system}, where a metallic nanomagnet is tunnel-coupled to two leads, one of which is also a magnet but with a fixed magnetization. The magnetization of the nanomagnet is dynamic and its angular dynamics are described by the LLGS equation. We treat the magnetization as a single macrospin with constant magnitude and include the Gilbert damping enhancement \cite{tserkovnyakEnhancedGilbertDamping2002} and a Slonczewski torque \cite{slonczewskiCurrentdrivenExcitationMagnetic1996,slonczewskiCurrentsTorquesPolarization2005,slonczewskiTheoryVoltagedrivenCurrent2007}.
	Applying a voltage over the system drives a spin-polarized current into the nanomagnet, which exerts a torque on its magnetization that can drive it into a precession \cite{slavinNonlinearAutoOscillatorTheory2009,kimChapterFourSpinTorque2012,chenSpinTorqueSpinHallNanoOscillators2016}. 
	
	Charge conservation requires that tunneling is the only way to change the charge of the nanomagnet and we therefore need to keep track of the number of electrons that tunnel into or out of the nanomagnet. This gives rise to a coupling between two degrees of freedom: (1) the nanomagnet's magnetization and (2) the nanomagnet's charge.
	On the one hand, the magnetization dynamics affects the charge degree of freedom by pumping a charge current through the system, since the left tunnel junction is magnetic \cite{tserkovnyakTunnelbarrierenhancedDcVoltage2008,ludwigBreakingCoulombBlockade2021a}. On the other hand, the charge degree of freedom affects the magnetization dynamics by altering the voltage drop across the magnetic tunnel junction, in turn altering the Slonczewski torque. As we will show, the emerging interplay gives rise to a self-induced torque in the magnetization dynamics. We emphasize that in this work we do not rederive any new results for the individual tunnel junctions; the self-induced torque emerges from charge conservation in the whole system. To demonstrate the relevance of the charge conservation in spin torque oscillators, we consider an in-plane magnetized nanopillar \cite{kiselevMicrowaveOscillationsNanomagnet2003} and show that accounting for the self-induced torque leads to qualitatively different predictions.

	The remainder of this work is organized as follows. We first describe the magnetization and charge dynamics and their coupling in Sec.~\ref{sec:coupled-dynamics}. In Sec.~\ref{sec:self-induced-torque} we show that this gives rise to a self-induced torque in a steady-state situation, which we show to be qualitatively and quantitatively important in the description of an in-plane magnetized nanopillar in Sec.~\ref{sec:exp-rel}. We end with a conclusion and discussion in Sec.~\ref{sec:discussion}.

	\section{Coupled dynamics of magnetization and charge}
	\label{sec:coupled-dynamics}
	The degrees of freedom of interest are the magnetization direction and the charge of the nanomagnet. For simplicity, we model the nanomagnet's magnetization by a single macrospin $\bm M$, which is justified when the nanomagnet is small enough. Similarly, we model the nanomagnet's charge by a single number $Q$, which describes the total charge on the nanomagnet. The coupled dynamics of magnetization and charge are then described by the Landau-Lifshitz-Gilbert-Slonczewski equation~\eqref{eq:LLGS} together with the continuity equation~\eqref{eq:conteq}.
	
	Assuming the magnetization's magnitude $M$ to be constant, the magnetization dynamics is described in terms of its direction $\bm m = \bm M / M$, which can also be specified with spherical coordinates $\bm m = ( \sin \theta \cos \phi, \sin \theta \sin \phi, \cos \theta )$, where we choose the left lead's fixed magnetization $\bm m_\mathrm{fix}$ as the $z$-axis. The equation of motion for $\bm m$ is the Landau-Lifshitz-Gilbert-Slonczewski (LLGS) equation \cite{slonczewskiCurrentdrivenExcitationMagnetic1996,ralphSpinTransferTorques2008},
	\begin{equation}
	\dot{\bm{m}} = - \gamma\, \bm m \times \bm H_\mathrm{eff}	 + \alpha(\theta)\, \bm{m} \times \dot{\bm{m}}  + \frac{\gamma}{M \mathcal{V}}\, \bm m \times \bm I_s \times \bm m\ .
			\label{eq:LLGS}
	\end{equation}
	Here, the first term describes the precession of the magnetization with a frequency $\omega$ around the effective magnetic field $\bm H_\mathrm{eff}$, where $\gamma$ is the gyromagnetic ratio. The second term, known as Gilbert damping \cite{gilbertPhenomenologicalTheoryDamping2004}, describes the relaxation of the magnetization towards an energetic minimum. The Gilbert-damping coefficient, $\alpha(\theta) = \alpha_0 + (\hbar^2 \gamma /4 e^2 M \mathcal{V})\, [\tilde g_l(\theta) + \tilde g_r]$, contains two terms corresponding to two sources of Gilbert damping: the first term, $\alpha_0$, accounts for internal Gilbert damping \cite{gilbertPhenomenologicalTheoryDamping2004}; the second term accounts for Gilbert-damping enhancement due to spin pumping into the leads \cite{tserkovnyakEnhancedGilbertDamping2002, brataasSpinBatteryOperated2002,tserkovnyakNonlocalMagnetizationDynamics2005,chudnovskiySpinTorqueShotNoise2008}, where $\mathcal{V}$ is the nanomagnet's volume and $\tilde g_l(\theta)$ and $\tilde g_r$ are the spin-flip conductances of the left and right junction respectively. The third term, known as Slonczewski torque, describes the torque arising due to the spin transfer from the electron system to the magnetization \cite{slonczewskiCurrentdrivenExcitationMagnetic1996,ralphSpinTransferTorques2008}.

	We assume strong internal relaxation in the nanomagnet's electron system, which means that it relaxes to equilibrium on time scales much shorter than the inverse tunneling rates. Electrons that enter the nanomagnet through one tunnel junction will therefore equilibrate with the other electrons in the nanomagnet before they leave again through the other tunnel junction; for a more detailed discussion, see App.~\ref{app:chemical}. In turn, both tunnel junctions are effectively independent. 
	The Slonczewski spin-transfer torque is then governed by 
	\begin{equation}
		\bm I_s  = -\frac{\hbar}{4e}\, g_l^s V_l\ \bm m_{\mathrm{fix}}, \label{eq:Is}
	\end{equation}
	where $g_l^s$ is the spin-flip conductance of the left junction, and $V_l$ is the voltage drop across the left junction \cite{slonczewskiTheoryVoltagedrivenCurrent2007,chudnovskiySpinTorqueShotNoise2008}.

	The charge dynamics are governed by charge conservation. The only way to change the nanomagnet's charge is by tunneling of electrons between the nanomagnet and its leads. Consequently, the charge dynamics are described by the continuity equation
	\begin{equation}
	\dot Q= I_c^{l} + I_c^{r}, \label{eq:conteq}
	\end{equation}
	where $I_c^{r}$ and $I_c^{l}$ are the charge currents flowing from respectively the right and left lead to the nanomagnet. The charge current through the right junction is simply given by the Ohmic relation
	\begin{equation}
		I_c^{r} = g_r V_r,
	\end{equation}
	where $g_r$ is the conductance of the right junction and $V_r$ is the voltage drop across the right junction \footnote{Strictly speaking, this is not an Ohmic current, because the power or energy loss in a tunnel junction is different from an Ohmic resistor.}. 
	
	In contrast to the right junction, the charge current across the left junction is spin-polarized, and thus the charge current is given by 
	\begin{equation}
		I_c^l = g_l(\theta) V_l + I_c^p,
	\end{equation}
	where  $g_l(\theta)$ is the conductance of the left junction, $V_l$ is the voltage drop across the left junction, and
	\begin{equation}
		I_c^p = \frac{\hbar}{4e}\,g_l^s\,  \bm m_\mathrm{fix} \cdot (\bm m \times \dot{ \bm m})\label{eq:Icp}
	\end{equation}
	is the pumping current contribution, which is the reciprocal of the Gilbert-damping enhancement, as it arises from the spin pumping into the leads combined with the spin filtering of the magnetic left lead \cite{tserkovnyakEnhancedGilbertDamping2002,tserkovnyakTunnelbarrierenhancedDcVoltage2008}. 
	
	The vector form, $I_c^p \propto \bm m_\mathrm{fix} \cdot (\bm m \times \dot{\bm m})$, holds for the specific geometry where the precession axis of the nanomagnet's magnetization is parallel to the fixed magnetization of the magnetic lead. However, we believe it to be valid in more general situations, since the cross product, $\bm m \times \dot{\bm{ m}}$, is related to the spin pumping which leads to the Gilbert-damping enhancement. The projection onto $\bm m_\mathrm{fix}$ enters due to the spin-filtering of the left lead, because of which the pumped spin current is accompanied by a pumped charge current. Therefore, in different geometries the same processes will occur, and thus the same vector form is to be expected. Finally we note that, since the left lead is magnetic, the conductance $g_l(\theta)$ also depends on the angle between the magnetizations of the nanomagnet and the left lead \cite{chudnovskiySpinTorqueShotNoise2008,sunMagnetoresistanceSpintransferTorque2008}.

	As the Landau-Lifshitz-Gilbert-Slonczewski equation~\eqref{eq:LLGS} and the continuity equation~\eqref{eq:conteq} are coupled to each other, an interplay emerges between the magnetization dynamics and the charge current flow: the magnetization dynamics affects the charge current flow via the pumping current $I_c^p$ {and} the charge current flow affects the magnetization dynamics via the Slonczewski spin-transfer torque $\bm I_s$.

	\section{Charge conservation gives rise to a self-induced torque}
	\label{sec:self-induced-torque}
	To investigate how the interplay with the charge dynamics affects the magnetization dynamics, we want to eliminate the charge degree of freedom. For that purpose, we focus on a steady-state situation for the charge (currents), $\dot Q=0$, which can be justified in two distinct ways: first, when we search for steady-state precessions of the magnetization---as in section \ref{sec:exp-rel}---the charge enters a steady-state as well. Second, the charge degree of freedom usually relaxes on much shorter time scales than the magnetization direction, such that the charge degree of freedom adjusts adiabatically, $\dot Q \approx 0$, to the magnetization dynamics  \cite{ludwigThermallyDrivenSpin2019}. 

	In a steady-state situation for the charge, the current flowing into the nanomagnet from the left lead is equal to the current flowing out of the nanomagnet into the right lead, $I_c^{l} = - I_c^{r}$, which follows immediately from the continuity equation \eqref{eq:conteq}. In turn, using Kirchhoff's voltage law $V_l - V_r = V$, we find the voltage drop across the left junction,
	\begin{equation}
	V_l = \frac{g_r}{g_l(\theta) + g_r} V - \frac{I_c^p}{g_l(\theta)+ g_r} . \label{eq:voltagedrop}
	\end{equation}
	The first term is the standard voltage drop for two resistors (or tunnel junctions) in series. The second term is more interesting: it is an additional voltage drop that arises due to the pumped charge current $I_c^p$, as given in Eq.~\eqref{eq:Icp}. Since this additional pumping-current-induced voltage drop is across the left junction with the magnetic lead, it also gives rise to an additional Slonczewski spin-transfer torque. Explicitly, using the voltage drop over the left junction, Eq.~\eqref{eq:voltagedrop} to find the Slonczweski spin-transfer-torque, Eq.~\eqref{eq:Is}, we obtain a modified LLGS equation~\eqref{eq:LLGS}, which now becomes
	\begin{multline}
			\dot{\bm{m}} = - \gamma\, \bm m \times \bm H_\mathrm{eff}	 + \alpha(\theta)\, \bm{m} \times \dot{\bm{m}} + \frac{\gamma}{M \mathcal{V}}\, \bm m \times  \bm I_s^V\!  \times \bm m \\
			 + \frac{\gamma}{M \mathcal{V}}\, \bm m \times \bm I_s^p \times \bm m\ .	 \label{eq:LLGS-final}
	\end{multline}
	where we have split up the spin-transfer torque in two contributions. Firstly, we obtain the standard Slonczewski spin-transfer torque induced by the voltage bias $V$ applied across the whole system,
	\begin{equation}
		\bm I_s^V = -\frac{\hbar}{4e}\frac{g_l^sg_r}{g_l(\theta) + g_r}\,V\, \bm m_{\mathrm{fix}}.
	\end{equation}
	It is simply proportional to the voltage bias applied over the two leads and has been obtained before by numerous authors \cite{slonczewskiTheoryVoltagedrivenCurrent2007,virtanenSpinPumpingTorque2017,chudnovskiySpinTorqueShotNoise2008}. However, we also obtain  
	\begin{equation}
		\bm I_s^p = \left(\frac{\hbar}{4e}\right)^2\frac{(g_l^s)^2}{g_l(\theta) + g_r} \left[\bm m_\mathrm{fix} \cdot (\bm m \times \dot{\bm m})\right]\ \bm m_\mathrm{fix} ,
		\label{eq:self-induced}
	\end{equation}
	which is our central result: a new self-induced torque. To be precise, it is a \emph{pumping-current induced} spin-transfer torque.
	Physically speaking, it originates from the pumping current $I_c^p$ driving electrons over a magnetic tunnel junction. If the precession axis of $\bm m$ is aligned with $\bm m_\mathrm{fix}$, it is simply proportional to the precession frequency $\omega$. Therefore---despite its origin---the self-induced torque effectively acts more similar to Gilbert damping than to the conventional voltage-bias induced spin-transfer torque. To be more precise, it will act effectively as an anti-Landau-Lifshitz damping, $\propto \bm m \times (\bm m_\mathrm{fix} \times \bm m)$, but with a prefactor that has a specific dependence of the magnetization dynamics, $\propto \bm m_\mathrm{fix} \cdot (\bm m \times \dot{\bm m})$. This specific dependence on the magnetization orientation and dynamics could be used to tell the self-induced torque apart from other spin-transfer torque terms and from the phenomenological nonlinear Gilbert damping.
	
	While a similar effect has been seen before in calculations far away from equilibrium \cite{ludwigStrongNonequilibriumEffects2017a,ludwigThermallyDrivenSpin2019,petrovicSpinChargePumping2018,bajpaiTimeretardedDampingMagnetic2019,bajpaiSpintronicsMeetsNonadiabatic2020,hammarTransientSpinDynamics2017}, we have shown here that those effects can also be interpreted as a consequence of charge conservation. We therefore expect this effect to be present independent of the strength of relaxation in the electron system, although it might change quantitatively with the internal relaxation rate.

	\section{Experimental relevance of the self-induced torque} \label{sec:exp-rel}
	In order to show that the self-induced torque is qualitatively and quantitatively relevant, we now consider an in-plane magnetized nanopillar \cite{kiselevMicrowaveOscillationsNanomagnet2003}. 
	The effective magnetic field for the nanomagnet is then $\bm H_{\mathrm{eff}} = H_0 \hat{\bm z}- 4\pi M_x \hat{\bm x}$, where the $\hat{z}$-direction is parallel to the magnetization of the left lead $\hat{z} = \bm m_\mathrm{fix}$ and $H_0$ is the applied external magnetic field. Then the frequency of the linear ferromagnetic resonance is $\omega_0=\gamma\sqrt{H_0(H_0 + 4\pi M)}$ and the critical voltage where the parallel alignment becomes unstable is $V_c=\gamma\,\alpha(0)(H_0 + 2\pi M) / \sigma_0$, where $\sigma_0\equiv \frac{\gamma}{M\mathcal V}\frac{\hbar}{4e}\frac{g_l^s}{g_l(0)+g_r}$ is the spin-polarization efficiency for a parallel alignment. Note that the critical voltage is independent of the self-induced torque, which only affects the dynamics in the auto-oscillation regime, where $V>V_c$.
	
	The charge conductances, $g_{l/r}(\theta)$, the spin-flip conductances, $\tilde g_{l/r}(\theta)$, and spin conductances, $g_{l/r}^s$, that characterize the tunnel junctions are not independent of each other. Instead, they are related to each other as \cite{sunMagnetoresistanceSpintransferTorque2008}
	\begin{gather}
		g_{l/r}(\theta) = \frac{1}{2}\left[g_{l/r}^{\mathrm P} + g_{l/r}^{\mathrm{AP}} + \left(g_{l/r}^{\mathrm P} - g_{l/r}^{\mathrm{AP}} \right) \cos\theta \right]\ , \ \ \ \ \nonumber \\
		\tilde g_{l/r}(\theta) = \frac{1}{2}\left[g_{l/r}^{\mathrm P} + g_{l/r}^{\mathrm{AP}} - \left( g_{l/r}^{\mathrm P} - g_{l/r}^{\mathrm{AP}} \right) \cos\theta \right]\ ,\\
		g_{l/r}^s = P^{-1} (g_{l/r}^{\mathrm P} - g_{l/r}^{\mathrm{AP}}), \nonumber
	\end{gather}
	where $g_{l/r}^{\mathrm{P}/\mathrm{AP}}$ are the conductances of the left/right tunnel junction with the magnetization parallel or anti-parallel to the fixed magnetization. Furthermore, $P=(\rho_m^\uparrow - \rho_m^\downarrow )/ (\rho_m^\uparrow + \rho_m^\downarrow) $ is the polarizing factor of the nanomagnet, where $\rho_m^\uparrow$ and $\rho_m^\downarrow$ are the nanomagnet's electron density of states for spin-up and spin-down respectively \cite{slonczewskiCurrentdrivenExcitationMagnetic1996}. Note that, since the right tunnel junction is nonmagnetic, we have $g_{r}^{\mathrm{P}}=g_{r}^{\mathrm{AP}}$ and it follows that $g_{r}(\theta) = \tilde g_{r}(\theta) =\vcentcolon g_r$ and $g_r^s = 0$. 
	
	\begin{figure}
		\includegraphics[width=\columnwidth]{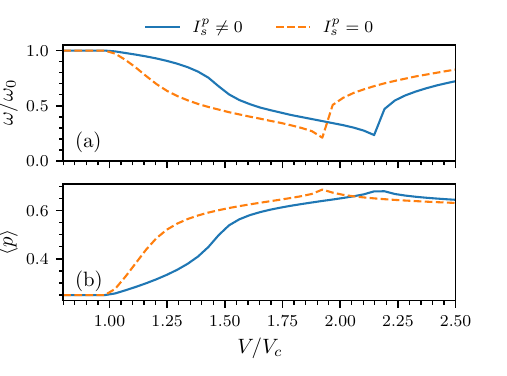}
		\caption{Numerical results, as obtained from the LLGS equation \eqref{eq:LLGS-final} with (solid) and artificially without (dashed) self-induced torque, are shown as a function of voltage for (a) the microwave frequency and (b) the average power $\langle p\rangle$. Results are expressed in units of the FMR frequency ($\omega_0$) and the critical voltage ($V_c$).  \label{fig:frequency-mz}}
	\end{figure}
	
	To demonstrate the relevance of the self-induced torque, we choose the polarization factor of Fe, $P=0.4$ \cite{meserveySpinpolarizedElectronTunneling1994}, and $g_l^{\mathrm P}/G_0=0.12$ and $g_l^{\mathrm{AP}}/G_0=0.05$, where $G_0=2e^2/h$ is the conductance quantum. The tunneling magnetoresistance (TMR) is then $\mathrm{TMR}=(R_{\mathrm{AP}} - R_{\mathrm{P}})/R_{\mathrm{P}}=70\%$ \footnote{Here, we defined $\mathrm{TMR}$ as in \textcite{sunMagnetoresistanceSpintransferTorque2008}.}. We choose the conductance of the right junction to be equal to the parallel conductance of the left junction, $g_r=g_l^{\mathrm P}$. Furthermore, we set $H_0/4\pi M=0.35$, $\hbar\gamma  / M\mathcal V=1$ and choose $\alpha_0$ such that $\alpha(0)=0.01$. The applied voltage $V$ is then variable, since in experiment it can also be directly controlled. We note that these values have been chosen to demonstrate the effect of the self-induced torque and do not necessarily correspond to a specific experiment.

	We numerically solve the LLGS Eq.~\eqref{eq:LLGS-final}, varying the applied voltage and we show the resulting microwave frequency and average power of the magnetization in Fig.~\ref{fig:frequency-mz} with (solid) and artificially without (dashed) the self-induced torque. Here $p=(1-m_z)/2$ is the power of the spin-torque oscillator \cite{slavinNonlinearAutoOscillatorTheory2009}. We give further details regarding the numerical simulation of the LLGS Eq.~\eqref{eq:LLGS-final} in App.~\ref{app:llgs}.

	The quantitative relevance of the self-induced torque for this specific set of parameters is clearly visible, since the solid and dashed lines in Fig.~\ref{fig:frequency-mz} are obviously different.
	The qualitative relevance of the self-induced torque can be seen from the onset of oscillations ($V \geq V_c$), where the self-induced torque causes the main oscillation frequency to plateau, Fig.~\ref{fig:frequency-mz} (solid), which is in qualitative agreement with experimental observations in similar systems \cite{kiselevMicrowaveOscillationsNanomagnet2003}. In contrast, if one only accounts for the conventional spin-transfer torque, Fig.~\ref{fig:frequency-mz} (dashed), the LLGS predicts a sharper drop of the main oscillation frequency at the onset of oscillations. A similar behavior is present for the average $z$-component, resulting in a plateau in the microwave power at the onset of oscillations.
	This nonlinear frequency shift at the onset of oscillation \footnote{Here, nonlinear refers to the fact that the precession frequency is proportional to $\langle p\rangle$.} is well known to be incompletely described by the conventional, or unmodified, LLGS equation \cite{tiberkevichNonlinearPhenomenologicalModel2007}. The inclusion of the self-induced torque corrects the qualitative behavior at the onset of oscillations. 
	We also observed that the self-induced torque extends the voltage-range over which auto-oscillations are allowed (not shown here). Here however the $\theta$-dependency of the Gilbert damping enhancement and the Slonczewski spin-transfer torque through the charge and spin-flip conductances $g_l(\theta)$ and $\tilde g_l(\theta)$ also plays an important role.

	\section{Discussion and conclusion}
	\label{sec:discussion}
	In this work, we have shown that charge conservation in spin torque oscillators can lead to a novel self-induced torque. Explicitly, we considered a magnetic double tunnel junction, Fig.~\ref{fig:system}, for which we showed that charge conservation leads to an interplay between magnetization and charge dynamics, which gave rise to a self-induced torque. This self-induced torque leads to important modifications in the precession power and frequency, especially at the onset of oscillations. It therefore offers an additional microscopic explanation for the experimental observation of a plateau at the onset of oscillations, complementing the phenomenological nonlinear Gilbert damping of Tiberkevich and Slavin \cite{tiberkevichNonlinearPhenomenologicalModel2007}.
	
	Due to their similar qualitative behavior, the self-induced torque could be easily mistaken as a contribution to nonlinear Gilbert damping. However, we do not suggest to replace the phenomenological nonlinear Gilbert damping of Tiberkevich and Slavin \cite{tiberkevichNonlinearPhenomenologicalModel2007} by the self-induced torque. It is likely that there are other effects, such as the magnon-electron and magnon-phonon interactions, that will lead to a proper nonlinear Gilbert damping. However, the self-induced torque and nonlinear Gilbert damping can be distinguished experimentally through their different angle dependence---as is clear from the vector form in Eq.~\eqref{eq:LLGS-final}.
	
	Self-induced torques similar to the one studied in this work, have previously been obtained. For example, Refs.~\cite{tserkovnyakDynamicStiffnessSpin2003,taniguchiCriticalCurrentSpintransfertorquedriven2008} found a backflow of spin current into the nanomagnet and associated self-induced torque by considering the spin accumulation at a normal metal in combination with imposing spin conservation. In addition, Ref.~\cite{chibaMagnetizationDampingNoncollinear2015} found a similar self-induced torque by imposing spin conservation in the interaction between two nanomagnets. However, these works required spin conservation in only a small part of the circuit, which is different from the approach we consider in our work: we impose charge conservation throughout the whole circuit. Because charge conservation cannot be easily violated, 
	we expect the self-induced torque obtained here to be an ubiquitous effect, which might also include the systems under study in Refs.~\cite{tserkovnyakDynamicStiffnessSpin2003,taniguchiCriticalCurrentSpintransfertorquedriven2008,chibaMagnetizationDampingNoncollinear2015}, although its qualitative and quantitative effects should be evaluated for each specific system. 
	
	Moreover, we expect the self-induced torque to be relevant beyond voltage-biased STOs. For example, it should be equally relevant in ferromagnetic resonance experiments, where a transverse oscillating magnetic field excites the macrospin \cite{tserkovnyakNonlocalMagnetizationDynamics2005}. Such experiments could be used to verify the existence of the self-induced torque proposed here and, simultaneously, to gain deeper insights into the nonlinear Gilbert damping as conceived by Tiberkevich and Slavin \cite{tiberkevichNonlinearPhenomenologicalModel2007}. Another important example, where the self-induced torque might play a crucial but underacknowledged role, is the magnetization switching in magnetic tunnel junctions \cite{bussmannSwitchingVerticalGiant1999,katineCurrentinducedRealignmentMagnetic2000}.

	Finally, we note that within spin torque oscillators the understanding of the linewidth has suffered from a similar problem as the precession frequency \cite{chudnovskiySpinTorqueShotNoise2008}. The often employed solution is to phenomenologically include the nonlinear Gilbert damping when describing the thermal fluctuations \cite{tiberkevichMicrowavePowerGenerated2007,kimGenerationLinewidthAutoOscillator2008,kimLineShapeDistortion2008a,slavinNonlinearAutoOscillatorTheory2009}. This solution, although effective in its description of the observed linewidth, does not offer a physical explanation of the nature of the noise. Following the results of this work, it would be of interest to further investigate the effects of charge conservation on the noise in spin torque oscillators. In particular, there will be an additional noise source coming from the altered voltage drop across the left junction as given in Eq.~\eqref{eq:voltagedrop}, which affects the spin shot noise associated with the discreteness of the spin passing through the left junction \cite{chudnovskiySpinTorqueShotNoise2008,virtanenSpinPumpingTorque2017}.

	\begin{acknowledgments}
	We thank A. Slavin and A. Shnirman for fruitful discussions.
	R.A.D. is member of the D-ITP consortium, a program of the Dutch Research Council (NWO) that is funded by the Dutch Ministry of Education, Culture and Science (OCW). This work is part of the research program Fluid Spintronics with project number 182.069, financed by the Dutch Research Council (NWO).
	\end{acknowledgments}

	\appendix
	
	\section{Electrochemical potentials}
	\label{app:chemical}
	To relate our simple approach to more elaborate calculations (for example Refs.~\cite{ludwigStrongNonequilibriumEffects2017a,ludwigThermallyDrivenSpin2019,petrovicSpinChargePumping2018,bajpaiTimeretardedDampingMagnetic2019,bajpaiSpintronicsMeetsNonadiabatic2020,hammarTransientSpinDynamics2017}), it is useful to connect the charge on the nanomagnet, $Q$, and the voltage drops across the tunnel junctions, $V_l$ and $V_r$, to the electrochemical potentials of the nanomagnet and the leads.
	
	The leads' electrochemical potentials are fixed by the setup (Fig. \ref{fig:system}) with a ground and voltage source. The electrochemical potential of the right lead $\mu_r$ serves as a reference point, as the right lead is connected to the ground. Thus, in the right lead, the electrochemical potential is identical with the chemical potential, as the electrical potential of the ground is (chosen to be) zero. The electrochemical potential of the left lead $\mu_l$ is then determined (in reference to $\mu_r$) by the applied voltage; explicitly, $\mu_l = \mu_r + eV$. In other words, the voltage that is applied across the whole system is determined by the difference between the leads' electrochemical potentials, $V = (\mu_l - \mu_r)/e$.
	
	The nanomagnet's electrochemical potential, $\mu_m$, is closely related to the nanomagnet's charge $Q = e \sum_\sigma \int \dd{\varepsilon} \rho_m^\sigma(\varepsilon)f_{m}(\varepsilon) + Q_b$, where $Q_b$ is the positive background charge, $\rho_m^\sigma(\varepsilon)$ is the nanomagnet's density of states for electrons with spin $\sigma$, and $f_{m}(\varepsilon) = (e^{(\varepsilon-\mu_m)/k_B T} + 1)^{-1}$ is the Fermi-Dirac distribution describing the electron distribution in the nanomagnet. Due to the close relation between $\mu_m$ and $Q$, we can use $\mu_m$ as a degree of freedom instead of $Q$ \footnote{For simplicity, we assume the temperature $T$ to be irrelevant for the currents, which is justified, for example, when the densities of states are approximately constant on the scale of $k_B T$.}. In direct analogue, an interplay now emerges between the magnetization dynamics and the nanomagnet's electrochemical potential.
	
	The electrochemical potential of the nanomagnet $\mu_m$ determines the voltage drops across the tunnel junctions through $V_{l/r} = (\mu_{l/r} - \mu_m)/e$. Using these relations for the voltage drops, and proceeding as in the main text, we can determine the steady-state electrochemical potential corresponding to equation \eqref{eq:voltagedrop}, 
	\begin{equation}
		\mu_{m} = \frac{g_l(\theta) \mu_l + g_r \mu_r}{g_l(\theta) + g_r} +   \frac{e I_c^p}{g_l(\theta) + g_r}. \label{eq: electrochemical potential}
	\end{equation}
	The electrochemical potential contains two terms: the first term is the standard result for double tunnel junctions with strong internal relaxation; the second term is a shift due to the pumping current. From equation \eqref{eq: electrochemical potential}, the interplay between the nanomagnet's electrochemical potential and its magnetization dynamics can be seen as follows. The magnetization dynamics affects the electrochemical potential $\mu_m$ via the pumping current $I_c^p$. The electrochemical potential $\mu_m$ directly affects the voltage drop across the left junction and, in turn, also the Slonczewski spin-transfer torque in equation \eqref{eq:LLGS}. As a result of this interplay, we find the LLGS equation \eqref{eq:LLGS-final} with a self-induced torque.
	
	Let us note that for $g_r \gg g_l(\theta), \tilde g_l(\theta), g_l^s$, we find $\mu_m \approx \mu_r$. Then, the voltage drop across the magnetic left junction becomes the same as the voltage applied across the whole system, $V_l \approx V$, and the self-induced torque becomes irrelevant. So, in this case, we can recover previous results for magnetic single tunnel junctions \cite{chudnovskiySpinTorqueShotNoise2008,virtanenSpinPumpingTorque2017}. However, when $g_r$ is roughly comparable to $g_l(\theta), \tilde g_l(\theta), g_l^s$, the self-induced torque is relevant. We expect this result to hold even if the tunnel junctions are replaced by direct contacts.
	
	Finally, however, let us also note that the electrochemical potential of the nanomagnet $\mu_m$ (and with it the voltage drops $V_l$ and $V_r$) is only well defined since we assume strong internal relaxation in the nanomagnet, which leads to an equilibrium Fermi-Dirac distribution for the electrons. Away from that local equilibrium, the nanomagnet's electrochemical potential is ill defined and the full electron distribution takes its role as degree of freedom. In turn, an interplay emerges between the magnetization dynamics and the electron distribution of the nanomagnet. While this interplay has been seen in the strong nonequilibrium case without internal relaxation \cite{ludwigStrongNonequilibriumEffects2017a,ludwigThermallyDrivenSpin2019,petrovicSpinChargePumping2018,bajpaiTimeretardedDampingMagnetic2019,bajpaiSpintronicsMeetsNonadiabatic2020,hammarTransientSpinDynamics2017}, the vast regime between the two limiting cases of strong and absent internal relaxation has---to the best of our knowledge---not yet been explored theoretically.
	
	\section{Numerical LLGS simulations}
	\label{app:llgs}
	The results as shown in Fig.~\ref{fig:frequency-mz} are obtained as follows. For every voltage we run a separate simulation of the LLGS equation~\eqref{eq:LLGS-final} numerically, starting from an initial condition where the macrospin is orientated along the $z$-axis, with a random small deviation. The simulations are then run for $4\pi \gamma M t=\num{5e4}$, in order to ensure a steady state precession has been reached. The precession frequency is then obtained from finding the highest peak in the Fourier transform of $m_x(t)+im_y(t)$ and the average $z$-component from averaging $m_z(t)$, where both are taken over the final $4\pi \gamma M t=\num{5e3}$. Since for the parameters chosen here we have that both $\alpha(\theta)\ll1$ and $\frac{\gamma}{M\mathcal V}|\bm I_s^p|\ll1$, the dynamics of the magnetization are governed primarily by $\dot{\bm m }= -\gamma\, \bm m \times \bm H_{\mathrm{eff}}$ and we can thus replace $\dot{\bm m }\rightarrow -\gamma\, \bm m \times \bm H_{\mathrm{eff}}$ on the right hand side of LLGS equation~\eqref{eq:LLGS-final} and in the self-induced torque, Eq.~\eqref{eq:self-induced}.

	\end{document}